\documentclass[twocolumn]{jpsj2} 

\title{$^{59}$Co-NMR Probe for Stepwise Magnetization and Magnetotransport in SrCo$_{6}$O$_{11}$ with Metallic Kagom\'e Layer and Triangular Lattice with Local Moments}

\author{Hidekazu Mukuda$^{1}$\thanks{E-mail address: mukuda@mp.es.osaka-u.ac.jp}, Yoshio Kitaoka$^{1}$, Shintaro Ishiwata$^{2,3}$, Takashi Saito$^{2}$, Yuichi Shimakawa$^{2}$, Hisatomo Harima$^{4}$ and Mikio Takano$^{2}$}

\inst{$^{1}$Department of Materials Engineering Science, Osaka University, Toyonaka, Osaka 560-8531 \\
$^2$Institute for Chemical Research, Kyoto University, Uji, Kyoto 611-0011\\
$^3$Department of Applied Physics, Waseda University, 3-4-1 Okubo, Shinjuku, Tokyo 169-8555\\
$^4$Department of Physics, Kobe University, Kobe 657-8501\\}

\abst{
We report on novel magnetic and electronic properties of SrCo$_6$O$_{11}$ that exhibits a unique stepwise  magnetization and its relevant magnetotransport phenomena investigated by the site-selective $^{59}$Co nuclear magnetic resonance (NMR) at zero and applied magnetic fields. This compound is composed of three Co sites in the unit cell, i.e., Co(1) in the metallic Kagom\'e layer, a Co(2) dimerized  pillar between the layers and Co(3) in the triangular lattice. Zero-field NMR spectra have revealed that large local moments at the Co(3) sites are magnetically ordered without any trace of bulk magnetization $M$ at zero field. The field-swept NMR spectra show that the internal hyperfine field at the Co(1) site is derived from fully polarized moments $M_s$ at the Co(3) sites in the "1"-plateau state at fields higher than 2.5 T, whereas it is partially cancelled out in the "1/3"-plateau state in which one-third of $M_s$ is induced at intermediate fields once a small field is applied. It has been clarified from a microscopic point of view that the local moments at Co(3) site undergo a field-induced ferrimagnetic ($\uparrow \uparrow \downarrow$)-to-ferromagnetic ($\uparrow \uparrow \uparrow$) transition. 
The Co(1) Kagom\'e layer and the dimerized pillar Co(2) site between the layers are of nonmagnetic origin, suggesting that the nearly quasi-2D metallic conductivity is dominated by nonmagnetic Co(1) and Co(2) sites. 
Consequently, unique magneto-transport phenomena observed in SrCo$_6$O$_{11}$ are demonstrated owing to the interaction between the conduction electrons at the  Co(1) and Co(2) sites and the local moments at Co(3) sites. }

\kword{cobalt oxide, Kagom\'e lattice, triangular lattice, magnetotransport, NMR}

\begin{document}
\maketitle


\section{Introduction}

A series of cobalt oxides, triggered by the observation of a large thermoelectric effect in Na$_{0.5}$CoO$_{2}$ \cite{Terasaki}, has attracted much attention since the discovery of superconductivity at $T_c\sim$5 K in Na$_{x}$CoO$_{2}$ $\cdot$ $y$H$_2$O ($x=0.35,y=1.3$) \cite{Takada}.  This is because it suggests a new route for searching high-$T_c$ superconductivity and also demonstrates the richness of the physics of layered transition metal oxides. This compound consists of two-dimensional CoO$_2$ layers separated by insulating blocks of Na$^{1+}$ and H$_2$O molecules, resembling a layered structure of copper oxide high-$T_c$ superconductors. Because of the octahedral crystal environment, Co$^{4+}$ for two-dimensional CoO$_2$ layers is in the low-spin $S=1/2$ state. The compound Na$_{0.35}$CoO$_2\cdot y$H$_2$O is considered to be a system in which 35\% electrons are doped to an $S=1/2$ triangular lattice. Insight from studies of this compound is expected to shed light on the mechanism of cuprate superconductors. Several experiments have, therefore, been conducted to investigate its physical properties and many theoretical proposals on the symmetry of superconductivity have been put forward. From an  other context, the frustration effects for magnetic interaction on transport properties have attracted much attention on triangular lattices \cite{Terasaki,Takada}, spinels\cite{spinel}, Kagom\'e lattices\cite{Kagome}, pyrochlore lattices \cite{Hanawa,Hiroi} and so on.

Recently, a novel cobalt oxide with the Kagom\'e lattice SrCo$_{6}$O$_{11}$ was synthesized using a high-pressure synthesis technique.\cite{Ishiwata} This compound consists of three different crystallographic cobalt sites (see Fig. \ref{Fig3}(a)): edge-sharing Co(1)O$_{6}$ octahedra forming a Kagom\'e lattice, face-sharing Co(2)$_2$O$_9$ dimers and trigonal Co(3)O$_5$ bipyramids between Kagom\'e layers.  Not only the Kagom\'e lattice of Co(1) but also Co(2) and Co(3) have own triangular sublattices, making a natural hybrid-frustration system. 
One of the unique features of SrCo$_{6}$O$_{11}$ is a stepwise increase in magnetization, i.e., an application of a low magnetic field ($H$) makes a magnetization increase from zero, giving rise to a "1/3" plateau of a fully polarized magnetization up to $H=2.5$ T,  and a further increase in $H$ leads to a fully polarized magnetization (called the "1" plateau) at  $H>2.5$ T.  Conductivity also shows a stepwise variation, suggesting a close relationship between magnetism and transport properties \cite{Ishiwata,Ishiwata_unpublished}.  The metallic conductivity on the Kagom\'e lattice reminds us of a metallic Na-Co-O triangular compound with frustrated electron spins.   

In this report, we unraveled the origin of novel magnetic and electronic properties of SrCo$_6$O$_{11}$  by means of site-selective $^{59}$Co-nuclear magnetic resonance (NMR) and nuclear quadrupole resonance (NQR), focusing on the close relationship between magnetism and transport properties. 

\section{Experimental}

Polycrystalline samples of SrCo$_6$O$_{11}$ were prepared by treating an appropriate precursor under high pressure as described elsewhere\cite{Ishiwata}. The samples were predominantly oriented to the $c$-axis in the presence of a magnetic field due to a strong Ising-type anisotropy along the $c$-axis. Thus, the NMR measurement in the field perpendicular to the $c$-axis was particularly performed using an oriented powder sample embedded in stycast 1266. Field-swept and frequency-swept NMR and  NQR spectra were obtained using a conventional phase-coherent-type NMR spectrometer. NMR spectroscopy is a powerful tool for clarifying the site dependences of the magnetic and electronic properties of this compound including three different Co sites. 

\section{Results and discussion}

\subsection{High-field NMR analyses ("1" plateau state)}

\begin{figure}[tb]
\begin{center}
\includegraphics[width=0.45\textwidth]{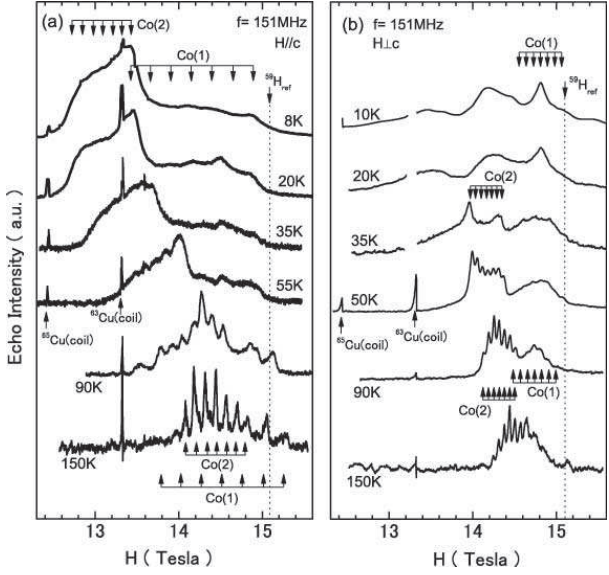}
\end{center}
\caption{Temperature ($T$) dependences of the field-swept $^{59}$Co-NMR spectra in high fields (a) parallel and (b) perpendicular to $c$-axis, corresponding to "1" plateau state. 
The spectra are composed of two superposed sets of resonance signals articulated by different nuclear quadrupolar splittings as indicated by the two sets of arrows. These are identified to arise from Co(1) and Co(2) sites(see text). At low temperatures, we deduce central peaks by assuming the same nuclear quadrupolar splittings obtained at high temperatures. }
\label{Fig1}
\end{figure}

Figures \ref{Fig1}(a) and \ref{Fig1}(b) show the temperature ($T$) dependences of the field-swept $^{59}$Co-NMR spectra in the fields parallel and perpendicular to the $c$-axis, corresponding to the "1" plateau state. 
The spectra are well articulated at high temperatures by a nuclear quadrupolar splitting, while it becomes broad with decreasing temperature due to the development of a short-range magnetic correlation. 
The spectra are composed of two superposed sets of resonance signals articulated by different nuclear quadrupolar splittings as displayed by the two sets of arrows in the figure: One has nuclear quadrupole frequencies of $\nu_{zz}(1)\approx$2.4 MHz and $\nu_{xx}(1)\sim$0.87 MHz and its in-plane asymmetric parameter $\eta(1)=|\nu_{xx}-\nu_{yy}|/\nu_{zz}\sim0.25$. The other has  $\nu_{zz}(2)\approx$1.24 MHz, $\nu_{xx}(2)\approx$0.67 MHz and $\eta(2)\sim0.08$. 
We assigned the former and latter to respective Co(1) and Co(2) sites from the following experimental results: (1) The Co(3) site was excluded since it had a large magnetic moment (see $\S$ 3.4). (2) Only hyperfine field for the former was cancelled out in the "1/3" state. It occurred only for the Co(1) site, taking account of the geometrical configuration and the origin of the  hyperfine field (see $\S$ 3.2). (3) The in-plane asymmetric parameter of nuclear quadrupole interaction, $\eta$, was expected to be larger for the Co(1) site forming the Kagom\'e lattice than for the Co(2) site, which was suggested by band calculation\cite{Harima}. 

\begin{figure}[tb]
\begin{center}
\includegraphics[width=0.45\textwidth]{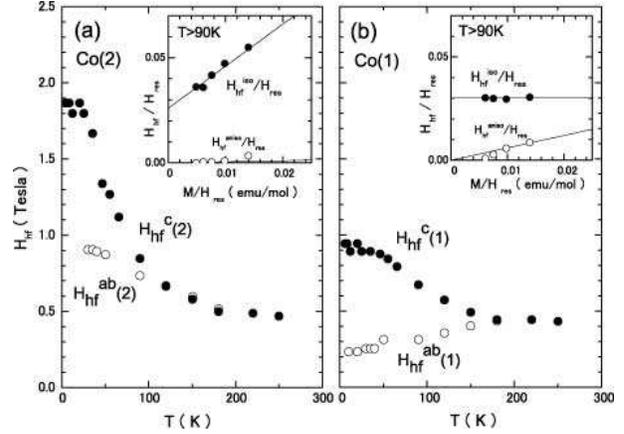}
\end{center}
\caption{$T$ dependences of $H_{\rm hf}$ at (a) Co(2) and (b) Co(1) sites. Insets show $H_{\rm hf}^{\rm iso,aniso}/H_{\rm res}$ against $M^{\rm iso,aniso}/H_{\rm res}$ with implicit parameter $T$. The isotropic hyperfine coupling constant is predominant for Co(2) sites, whereas the anisotropic one is predominant for the Co(1) site, suggesting that the former and latter are derived from the transferred field and dipole field from the magnetic moment of the Co(3) site, respectively. }
\label{Fig2}
\end{figure}

The hyperfine field is obtained using  $H_{\rm hf}^{\rm c,ab}=^{59}H_{\rm ref}-H_{\rm res}^{\rm c,ab}$, where $^{59}H_{\rm ref}=f/^{59}\gamma_{\rm n}$ is a reference field corresponding to $H_{\rm hf}^{\rm c,ab}=0$, and $H_{\rm res}^{\rm c,ab}$ is deduced from the resonance field of the central peak when applying the fields parallel and perpendicular to the $c$-axis, respectively. 
As shown in Figs. \ref{Fig2}(a) and \ref{Fig2}(b), $H_{\rm hf}^{\rm c,ab}$ shows a strong $T$ dependence, corresponding to that of magnetic susceptibility. 
The decrease in resistivity around 50 K observed at $H=5$ T \cite{Ishiwata_unpublished} is associated with the suppression of spin-spin scattering due to a possible onset of some ferromagnetic ordering. 
To analyze the origin of the hyperfine field, we extract the isotropic and anisotropic parts of $H_{\rm hf}$ given by $H_{\rm hf}^{\rm iso}=(H_{\rm hf}^{\rm c}+2H_{\rm hf}^{\rm ab})/3$ and $H_{\rm hf}^{\rm aniso}=(H_{\rm hf}^{\rm c}-H_{\rm hf}^{\rm ab})/3$, respectively.
Generally, the hyperfine field at Co sites involves orbital and spins contributions, and is expressed by 
$H_{\rm hf} = K_{\rm orb} H_{\rm res} + A_{\rm s} M $, where $K_{\rm orb}$, $A_{\rm s}$ and $M$ are the orbital component of the Knight shift, the spin component of the hyperfine coupling constant and the bulk magnetization, respectively. 
As shown in the insets of Figs. \ref{Fig2}(a) and \ref{Fig2}(b), $H_{\rm hf}^{\rm iso,aniso}/H_{\rm res}$ are plotted against $M^{\rm iso,aniso}/H_{\rm res}$ \cite{M/H} with an implicit parameter of $T$. 
From a linear relation in the paramagnetic region at high temperatures, $K_{\rm orb}(2)$ and $A_{\rm s}^{\rm iso}(2)(A_{\rm s}^{\rm aniso}(2)$) for the Co(2) site are estimated to be $\sim$2.7\% and $\sim$11.2(0.3) kOe/$\mu_{\rm B}$ per formula unit, respectively. 
This suggests that the isotropic hyperfine interaction is predominant in the case of the Co(2) site. 
The small and positive $A_{\rm s}^{\rm iso}(2)$ value for the Co(2) site indicates the presence of the transferred hyperfine field from neighboring magnetically ordered moments at the Co(3) sites through Co(2)$^{\rm 4s}$$\Leftarrow$O$^{2p}$$\Leftarrow$Co(3)$^{3d\uparrow}$ covalent bonding. 
Here, it is expected that the local moment at the bipyramidal Co(3) site is responsible for the magnetic order in this compound, as will be shown later.
The recent magnetization measurement suggests that the saturated moment is approximately 4$\mu_{\rm B}$ corresponding to that Co(3) is a high-spin Co$^{4+}$ configuration with $S=2$ \cite{Saito}. 

\begin{figure}[tb]
\begin{center}
\includegraphics[width=0.45\textwidth]{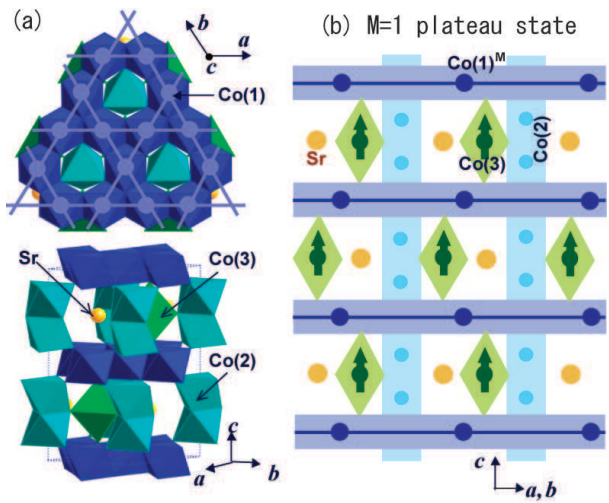}
\end{center}
\caption{(a) Crystal structure of SrCo$_6$O$_{11}$. (b) Schematics of cross-sectional view of plane including Co(1), Co(2) and Co(3). Spectrum analyses revealed that Co(1) and Co(2) sites are affected by the respective transferred field and dipole field from ferromagnetically ordered moments at the Co(3) site for the "1" plateau state. }
\label{Fig3}
\end{figure}

By the same manner, $K_{\rm orb}(1)$ and $A_{\rm s}^{\rm iso}(1)(A_{\rm s}^{\rm aniso}(1)$) for the Co(1) site are estimated to be $\sim$3\% and $\sim$0(3.4) kOe/$\mu_{\rm B}$ per a formula unit, respectively. 
In this case, the anisotropic hyperfine interaction is dominant although the isotropic part is negligible. 
The anisotropic part of the hyperfine field usually comes from the dipole interaction from neighboring local moments. 
Figure \ref{Fig3}(b) shows the schematics of the cross-sectional view in the plane including the Co(1), Co(2) and Co(3) sites. 
All of the Co(1) sites should be affected by the dipole field from the Co(3) site, which is calculated to be $\sim$ 0.8 kOe/$\mu_{\rm B}$, taking the ferromagnetically aligned neighboring Co(3) into account. 
This value is slightly smaller than the $A_{\rm s}^{\rm aniso}(1)$ estimated experimentally, suggesting that the $A_{\rm s}^{\rm aniso}(1)$ includes the pseudodipole interaction in addition to the classical dipole field from the ferromagnetically aligned local moment at the Co(3) site.

\subsection{Low-field NMR analyses ("1/3" plateau state)}

\begin{figure}[tb]
\begin{center}
\includegraphics[width=0.5\textwidth]{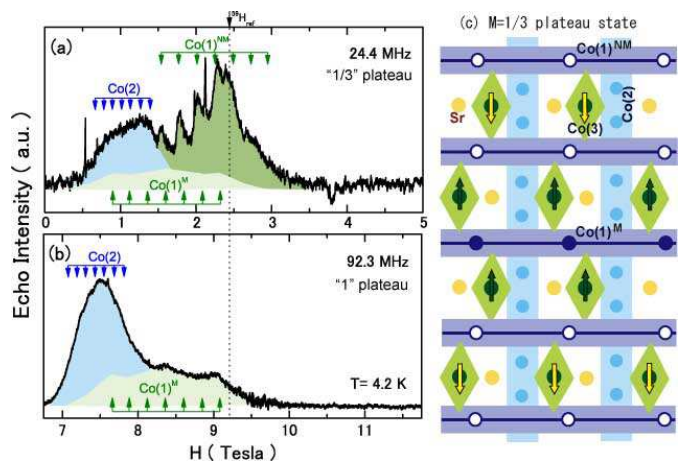}
\end{center}
\caption{Co-NMR spectra of 4.2K obtained at (a) "1/3" ($H \le 2.5$ T ) and (b) "1" plateaus ($H \gg 2.5$ T ). The dotted vertical lines in both spectra commonly point to $H_{\rm hf}=0$. (a)The spectrum in the "1/3" plateau state consists of a broad peak largely shifted from $H_{\rm hf}=0$ arising from the Co(2) site and an additional well-articulated peak around $H_{\rm hf}=0$ arising from the \textit{nonmagnetic} Co(1)$^{\rm NM}$ site.  (c) The Co(1)$^{\rm M}$ and Co(1)$^{\rm NM}$ sites (denoted by closed and open circles) are surrounded by ferromagnetically and antiferromagnetically coupled moments at two neighboring Co(3) sites, respectively. Note that a hyperfine field is canceled out at the Co(1) site only for the "1/3" plateau state. }
\label{Fig4}
\end{figure}
The magnetization measurements revealed a stepwise increase with increasing $H$, i.e., one third of fully polarized magnetization $M_{\rm s}$ is induced at an intermediate state (called the "1/3" plateau state) once a tiny field is applied. When exceeding $H=$2.5 T, however, the moment at the Co(3) site is fully polarized, leading to a saturation of magnetization $M_{\rm s}$ called the "1" plateau state. 

Figures \ref{Fig4}(a) and \ref{Fig4}(b) show the NMR spectra of 4.2 K obtained at 24.4MHz ($H \le 2.5$ T) and 92.3MHz ($H \gg 2.5$ T)  corresponding to fields that exhibit "1/3 and "1" plateaus, respectively. In the "1" plateau state (Fig. \ref{Fig4}(b)), the spectra for the Co(2) and Co(1) sites are largely affected by the respective transferred hyperfine field and dipole field from the Co(3) site, as discussed in $\S$ 3.1.  
By contrast, the spectrum in the "1/3" plateau state (Fig. \ref{Fig4}(a)) was markedly different from that in the "1" plateau state: It consists of a broad peak largely shifted from $H_{\rm hf}=0$ arising from the Co(2) site and an additional well-articulated peak around $H_{\rm hf}=0$ arising from the \textit{nonmagnetic} Co(1)$^{\rm NM}$ site since its spectrum is articulated by the nuclear quadrupole frequency at the Co(1) site ($\nu_{\rm zz}(1)$). The presence of Co(1)$^{\rm NM}$ in the "1/3" plateau state suggests that the hyperfine field derived from the magnetic moments at the Co(3) site is cancelled out because the magnetic moments Co(3) in the vicinity of Co(1)$^{\rm NM}$ are antiferromagnetically coupled in the "1/3" plateau state. 
Recently, Saito \textit{et al.} have found the magnetic superlattice reflections ($a\times a\times c$) in the  "1/3" plateau state by a neutron diffraction study and determined a ferrimagnetic structure with ($\uparrow \uparrow \downarrow$) along the $c$-axis\cite{Saito}. 
Therefore, it is plausible that such a ferrimagnetic structure is realized by ferromagnetically aligned Co(3) local moments within the triangular lattice, as displayed in Fig. \ref{Fig4}(c). 
In the "1/3" state, we can speculate the presence of \textit{magnetic} Co(1)$^{\rm M}$ sites between ferromagnetically coupled Co(3) moments, considering the magnetic structure shown in Fig. \ref{Fig4}(c). 
Namely, two-thirds of Co(1) sites become nonmagnetic and the remaining one-third maintains magnetic state. 
Although the resonance signal of Co(1)$^{\rm M}$ sites cannot be observed clearly in the spectrum, its spectral shape may be assumed to be the same as that observed in the "1" plateau state as shown in Fig. \ref{Fig4}(a).  Here, we note that the hyperfine field for the Co(2) site is not cancelled out in this structural configuration. 

Therefore, it is concluded that Co(3) sites in the triangular lattice undergo the local moment at the Co(3) aligned ferromagnetically in the triangular plane with the modulated magnetic structure of ($\uparrow \uparrow \downarrow$) along the $c$-axis in the "1/3" plateau state to the ferromagnetic ($\uparrow \uparrow \uparrow$) transition in the "1" plateau state. The stepwise increase in $M$ microscopically corroborates the marked change in Co(1)-NMR spectral shape at fields below and above $H=2.5$ T.  

\subsection{Transition between "1/3" and "1" plateaus}

\begin{figure}[tb]
\begin{center}
\includegraphics[width=0.4\textwidth]{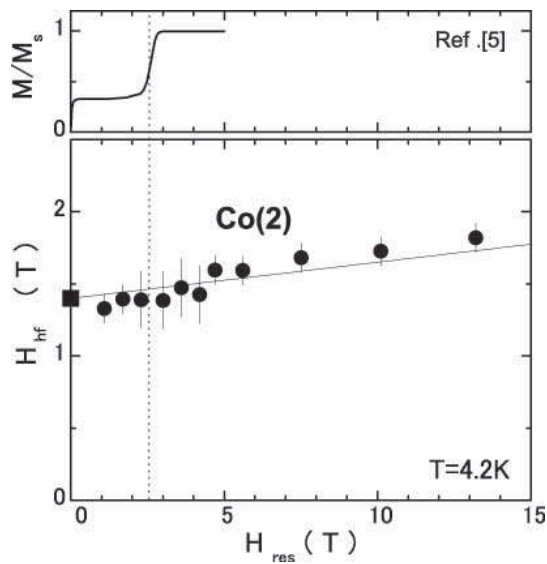}
\end{center}
\caption{Field dependence of hyperfine field for Co(2) site. Unexpectedly, the local magnetization at the Co(2) site $H_{\rm hf}(2)$ does not exhibit such a marked variation around 2.5 T where a prominent stepwise magnetic transition was found in the uniform magnetization (upper figure\cite{Ishiwata}). It is also evident that there is no spontaneous magnetization at the Co(2)$^{\rm M}$ site in both the "1/3" and "1" plateau states, and that  stepwise magnetism in this compound should be attributed to the configuration of the magnetic structure of the Co(3) site, not to the size of the magnetic moment nor the contribution of the other Co sites except Co(3) sites. The closed square is the internal field observed at zero external field in Fig.\ref{Fig6}.}
\label{Fig5}
\end{figure}

To gain further insight into the transition between the "1/3" and "1" plateaus, we investigated the field dependence of the hyperfine field $H_{\rm hf}$(2) at the Co(2) site at 4.2 K where the local magnetic moment of Co(3) saturates. As seen in Figs. \ref{Fig4}(a) and \ref{Fig4}(b), $H_{\rm hf}$(2) determined from $^{59}H_{\rm ref}-H_{\rm res}$ varies as a function of $H$ at a fixed temperature. Unexpectedly, the local magnetism at the Co(2) site does not exhibit any drastic variation around $H=2.5$ T, irrespective of the stepwise increase in $M$\cite{Ishiwata}, as indicated in Fig. \ref{Fig5}. As discussed in $\S$ 3.1, $H_{\rm hf}$(2) is predominantly derived from the transferred field  from the Co(3) site in addition to the orbital contribution induced by the external field. The observed $H$ dependence of $H_{\rm hf}$(2) is well reproduced by the calculated line evaluated from the relation $H_{\rm hf}=K_{\rm orb}(2)H_{\rm res}+ H_{\rm s}$, where $K_{\rm orb}(2)\sim 2.7$\% and $H_{\rm s}$ is a spin component of the hyperfine field. This indicates that the $H$-dependent component mainly stems from the orbital contribution $\sim K_{\rm orb}(2)H_{\rm res}$, and that the spin contribution $H_{\rm s}$ is nearly $H$-independent. Therefore, the local magnetic state of Co(3) does not change above and below 2.5 T, keeping a constant saturated moment $M_{\rm s}=4\mu_{\rm B}$. Note that the presence of $H_{\rm hf}$(2) of 1.4 T even at zero field gives evidence that the system magnetically orders at $H=0$, as estimated from the zero-field NMR spectra shown in the next section. It is also evident that there exists no spontaneous magnetization at the Co(2) site in both the "1/3" and "1" plateau states, and that the stepwise magnetism in this compound should be attributed to the configuration of the magnetic structure of the Co(3) site and not to the size of the magnetic moment nor the contribution from the other Co sites except Co(3) sites. Thus, both Co(1) and Co(2) sites are of nonmagnetic origin, which is consistent with the results of the band calculation elucidating that a nearly quasi-2D metallic conductivity is dominated by nonmagnetic Co(1) and Co(2) sites\cite{Ishiwata_unpublished}. 

\subsection{Possible magnetic ground state at zero external field; transition between "0" and "1/3" states}

\begin{figure}[tb]
\begin{center}
\includegraphics[width=0.4\textwidth]{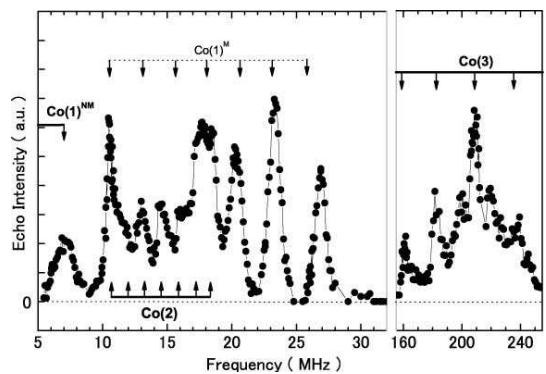}
\end{center}
\caption{Frequency-swept zero-field NMR/NQR spectrum at 1.6K. The resonance signal in the high-frequency region around 209 MHz is assigned to be Co(3) because of the largest NQR splitting induced by $\nu_{\rm Q}(3)\sim$26 MHz.  The spectra from 10 MHz to 18 MHz are identified as the magnetic Co(2) with  the transferred hyperfine field of 1.4 T. The remaining peaks found at 10$\sim$27 MHz may have arisen from the magnetic Co(1)$^{\rm M}$ site, which is reproduced by $\nu_{zz}(1)\approx$ 2.4 MHz at an internal field of 1.8 T. }
\label{Fig6}
\end{figure}

Each magnetic characteristic at the three cobalt sites is identified from the observation of the zero-field(ZF)-NMR / NQR spectrum at 1.6 K, as shown in Fig. \ref{Fig6}. The ZF-NMR spectrum in the high-frequency region around 209 MHz is considered to arise from the Co(3) site. This is because the largest NQR frequency $\nu_{\rm zz}(3)\sim 26$ MHz among the three Co sites is due to the local trigonal symmetry for the bipyramidal structure at the Co(3) site. Here, $\nu_{\rm zz}(3)$ is evaluated from the splitting of the resonance peaks indicated by arrows in the figure. The hyperfine field $H_{\rm hf}(3)$ for Co(3) site is evaluated to be about -20.9 T from the relation $f_{\rm res}/^{59}\gamma_{\rm n}$. The hyperfine coupling constant $A_{\rm hf}(3)$ at the Co(3) site is estimated to be -5 T/$\mu_{\rm B}$ by assuming the saturated moment $M_s\sim4\mu_{\rm B}$. Here, the sign of $A_{\rm hf}(3)$ is expected to be negative via the $3d$ core polarization effect that used to be predominant for magnetic transition metal ions. 

When noting that the ZF-NMR spectrum at the Co(3) site exhibits no apparent change even though the small amount of field is applied to stabilize the "1/3" plateau state, the local magnetic nature at the Co(3) site does not change in going from the "0" state to the "1/3" state with the stepwise increase in $M$. 

The NQR and ZF-NMR spectra observed below 30 MHz in Fig. \ref{Fig6} are quite complicated because of the mixture of nonmagnetic Co(1)$^{\rm NM}$, the magnetic Co(1)$^{\rm M}$ and Co(2) sites. The spectrum from 10 MHz to 18 MHz is identified as the Co(2) site because the spectrum is articulated by the nuclear quadrupole splitting of the Co(2) site $\nu_{zz}(2)$,  and the internal field is estimated to be 1.4 T comparable to the field extrapolated value in Fig. \ref{Fig5}. As indicated in the figure, the hyperfine field for the Co(2) site exhibits no marked change in the stepwise increase from the "0" state to the "1" state through the "1/3" state, since its hyperfine field is transferred from the local moments at the Co(3) site. This result suggests that the local moments at the Co(3) sites of $M_s=4\mu_B$ exist irrespective of the stepwise magnetization from the "0" to "1" states.
The NQR spectrum of Co(1)$^{\rm NM}$ is detected below 7 MHz corresponding to the $\pm5/2 \leftrightarrow \pm 7/2$ transition. The remaining peaks from 10 to 27 MHz are assumed to be the NMR spectrum for the Co(1)$^{\rm M}$ site, because the peaks are approximately separated by $\nu_{zz}(1)\sim$ 2.4 MHz. In this case, the internal fields at Co(1) sites is 1.8 T at 1.6 K, which is slightly larger than the expected value extrapolated in the  "1/3" plateau state. Further measurements, such as those of the field dependence of this NQR spectrum and the nuclear spin-lattice relaxation rate measurement, will give a firm evidence of the site assignment in the ZF-NMR spectrum and the magnetic structure at zero field.

Next, we address a possible nature of the magnetically ordered state in this compound, which is inferred from the present ZF-NMR results. The ground state of frustrated spins on a triangular lattice with and without the magnetic field has attracted intensive interest over the years. The spin $S=1/2$ antiferromagnetic Ising-like Heisenberg model on a triangular lattice is expected to have no uniform magnetization at $H=0$ and shows a 1/3 magnetization plateau \cite{Nishimori}.
For SrCo$_6$O$_{11}$, however, the Co(1) site in the Kagom\'e layer and the dimerized pillar Co(2) site between the layers are of nonmagnetic origin, suggesting that a nearly quasi-2D metallic conductivity is dominated by  nonmagnetic Co(1) and Co(2) sites, as supported by the band calculation \cite{Ishiwata_unpublished}.
The unique stepwise increase in the $M$ of SrCo$_6$O$_{11}$ originates from the large local moment at the Co(3) site. Therefore, $M$=0 at $H$=0 suggests that the $M$=0 state is realized by the cancellation of macroscopic magnetization due to the domain structures of $(\uparrow\uparrow\downarrow)$ and $(\downarrow\downarrow\uparrow)$ along the $c$-axis, which is formed by the ferromagnetically ordered Co(3) triangular plane. 
Further neutron scattering experiment is desired for determining the magnetic structure at $H$=0 in SrCo$_6$O$_{11}$.

\section{Conclusion}

A site-selective $^{59}$Co-NMR study of SrCo$_6$O$_{11}$ has revealed that a unique stepwise increase in magnetization stems from the rearrangement of the local moments at the trigonal bipyramidal Co(3) site. The Co(1) Kagom\'e layer and the dimerized  pillar Co(2) site between the layers are of nonmagnetic origin, suggesting that the nearly quasi-2D metallic conductivity is dominated by nonmagnetic Co(1) and Co(2) sites. The transferred hyperfine field is cancelled in two-thirds of Co(1) sites in the "1/3" plateau state, whereas all  Co(1) sites in the "1" plateau state are affected by ferromagnetically ordered moments at Co(3) sites. Co(3) sites in the triangular lattice undergo a local moment aligned ferromagnetically in the triangular plane with the ferrimagnetic structure  ($\uparrow \uparrow \downarrow$) along the $c$-axis in the "1/3" plateau state, which is consistent with the evidence obtained from the recent neutron diffraction experiment. By applying the field, SrCo$_6$O$_{11}$ exhibits a transition to the ferromagnetic ($\uparrow \uparrow \uparrow$) state corresponding to the "1" plateau state. The large local moments at Co(3) sites are demonstrated to be in a magnetically ordered state even when $M$=0 at $H$=0, suggesting the cancellation of the bulk magnetization due to a domain structure of the ferrimagnetic state. Consequently the unique magnetotransport phenomena observed in SrCo$_6$O$_{11}$ are demonstrated due to the interaction between conduction electrons at Co(1) and Co(2) sites and local moments at Co(3) sites.

The authors are grateful to N. Nagaosa, F. Ishii, H. Harima and G.-q. Zheng for fruitful discussion and comments, and T. Ohara for experimental assistance. This work was supported by a Grant-in-Aid for Creative Scientific Research (15GS0213) from the Ministry of Education, Culture, Sports, Science and Technology (MEXT) and the 21st Century COE Program (G18) by Japan Society of the Promotion of Science (JSPS). 



\begin{thebibliography}{99} 

\bibitem{Terasaki} I. Terasaki, Y. Sasago, and K. Uchinokura: Phys. Rev. B \textbf{56} (1997) R12685.
\bibitem{Takada} K. Takada, H. Sakurai, E. Takayama-Muromachi, F. Izumi, R. A. Dilanian, and T. Sasaki: Nature \textbf{422} (2003) 53.
\bibitem{spinel} S. Kondo, D. C. Johnston, C. A. Swenson, F. Borsa, A. V. Mahajan, L. L. Miller, T. Gu, A. I. Goldman, M. B. Maple, D. A. Gajewski, E. J. Freeman, N. R. Dilley, R. P. Dickey, J. Merrin, K. Kojima, G. M. Luke, Y. J. Uemura, O. Chmaissem, and J. D. Jorgensen: Phys. Rev. Lett. 78 (1997) 3729.
\bibitem{Kagome}C. Broholm, G. Aeppli, G. P. Espinosa, and A. S. Cooper: Phys. Rev. Lett. 65 (1990) 3173. 
\bibitem{Hanawa} M. Hanawa, Y. Muraoka, T. Tayama, T. Sakakibara, J. Yamaura, and Z. Hiroi: Phys. Rev. Lett. \textbf{87} (2001) 187001.
\bibitem{Hiroi} Z. Hiroi, S. Yonezawa, and Y. Muraoka: J. Phys. Soc. Jpn. \textbf{73} (2004) 1651.
\bibitem{Ishiwata}S. Ishiwata, D. Wang, T. Saito, and M. Takano: Chem. Mater. \textbf{17} (2005) 2789.
\bibitem{Ishiwata_unpublished} S. Ishiwata {\it et al.}: in preparation. 
\bibitem{M/H} $M^{\rm c}/H_{\rm res}$ data measured at 5T using SQUID magnetometer are used. In the insets of Figs. \ref{Fig2}(a) and \ref{Fig2}(b), $H_{\rm hf}^{\rm iso}/H_{\rm res}$ and $H_{\rm hf}^{\rm aniso}/H_{\rm res}$ are plotted against $M^{\rm iso}$ and $M^{\rm aniso}$ given by $(M^{\rm c}+2M^{\rm ab})/3$ and $(M^{\rm c}-M^{\rm ab})/3$, respectively.
\bibitem{Saito} T. Saito, A. Williams, J. P. Attfield, T. Wuernisha, T. Kamiyama, S. Ishiwata, Y. Takeda, Y. Shimakawa and M. Takano: preprint. 
\bibitem{Nishimori} H. Nishimori and S. Miyashita: J. Phys. Soc. Jpn. \textbf{55} (1986) 4448.
\bibitem{Harima} H. Harima: private communication.

\end{thebibliography}
\end{document}